\documentclass[10pt,conference]{IEEEtran}
\usepackage{graphicx} 
\usepackage{subcaption}
\usepackage{xcolor}
\usepackage{color}
\usepackage[style=ieee,sortcites=true]{biblatex} 
\usepackage{siunitx}
\usepackage{hyperref}
\usepackage{amsmath}
\usepackage{mathtools}
\usepackage{wrapfig}
 \usepackage{booktabs}
\DeclarePairedDelimiter{\ceil}{\lceil}{\rceil}

\title{Quantum Annealing Hyperparameter Analysis for Optimal Sensor Placement in Production Environments}
\author{
    \IEEEauthorblockN{
    Nico Kraus\IEEEauthorrefmark{1}, 
    Marvin Erdmann\IEEEauthorrefmark{2}, 
    Alexander Kuzmany\IEEEauthorrefmark{1},
    Daniel Porawski\IEEEauthorrefmark{1},
    Jonas Stein\IEEEauthorrefmark{3}}
    \IEEEauthorblockA{\IEEEauthorrefmark{1}Aqarios GmbH, Munich, Germany}
    \IEEEauthorblockA{\IEEEauthorrefmark{2}BMW Group, Munich, Germany}
    \IEEEauthorblockA{\IEEEauthorrefmark{3}LMU Munich, Germany}
}

\date{\today}

\begin{document}

\maketitle
\begin{abstract}

To increase efficiency in automotive manufacturing, newly produced vehicles can move autonomously from the production line to the distribution area. This requires an optimal placement of sensors to ensure full coverage while minimizing the number of sensors used. The underlying optimization problem poses a computational challenge due to its large-scale nature.
Currently, classical solvers rely on heuristics, often yielding non-optimal solutions for large instances, resulting in suboptimal sensor distributions and increased operational costs.



We explore quantum computing methods that may outperform classical heuristics in the future. We implemented quantum annealing with D-Wave, transforming the problem into a quadratic unconstrained binary optimization formulation with one-hot and binary encoding. Hyperparameters like the penalty terms and the annealing time are optimized and the results are compared with default parameter settings. 

Our results demonstrate that quantum annealing is capable of solving instances derived from real-world scenarios. Through the use of decomposition techniques, we are able to scale the problem size further, bringing it closer to practical, industrial applicability. Through this work, we provide key insights into the importance of quantum annealing parametrization, demonstrating how quantum computing could contribute to cost-efficient, large-scale optimization problems once the hardware matures.

\end{abstract}
\section{Introduction}
\label{sec:intro}

Quantum computing represents a transformative paradigm shift in computational capabilities, offering unprecedented potential for solving complex problems across various domains~\cite{bayerstadler_industry_2021}. Among its many applications, quantum computing could help solve highly complex, extremely large-scale optimization problems faster and better than classical methods when the hardware matures.

This paper evaluates the potential of using quantum annealing to optimize the placement and configuration of LiDAR sensors in automotive production environments to enable vehicles to move autonomously from the end of the production line to their respective distribution areas~\cite{bmw_2024}. 
In these environments, vehicles must navigate mixed traffic, including pedestrian zones and intersections, making precise and efficient sensor placement critical. This use case addresses the challenge of maximizing coverage while minimizing the number of sensors, a task that is inherently complex due to the vast number of potential sensor configurations (spatial and angular in sufficiently high resolution). Traditional optimization methods, while effective, often fall short in handling the extensive solution spaces required for real-world applications.

Quantum computing offers a promising alternative, particularly through faster runtimes compared to classical methods \cite{shor}. Quantum annealing offers a compelling alternative for optimization problems, enabling more efficient exploration of large configuration spaces \cite{Kadowaki_1998, bauza2024}. This could lead to significant cost savings, given the high long-term investment per sensor, and improved resolution in sensor coverage—allowing detection of smaller obstacles and enhanced autonomous navigation.

Our work could serve as a basis for further exploration into the application of quantum computing for optimizing sensor configurations in various contexts, such as managing intersections in public traffic~\cite{10237193}. As the prevalence of autonomous vehicles increases, 
advanced optimization strategies for sensor deployment will be essential to ensure safety and operational efficiency in dense and dynamic environments.

This paper makes several significant contributions to the field of quantum computing for sensor configuration optimization. First, we formulate a problem model that captures the key features of the optimization problem, is compatible with quantum annealing devices, and can be evaluated on scenarios involving up to 4109 qubits. 
Second, we perform a thorough analysis of quantum annealing hyperparameters, demonstrating how tuning these can significantly improve performance over default settings in problem instances ranging from small toy problems to real-world inspired scenarios closer to industrial applicability. Lastly, we highlight the effectiveness of problem-specific decomposition techniques, showcasing their utility in enhancing solution quality for large-scale optimization problems.



\section{Background}
\label{sec:background}

\subsection{Problem Formulation}

The sensor placement problem is defined by a bipartite graph with vertices for each potential LiDAR sensor position ($l \in V_l$) and street points ($s \in V_s$). If a LiDAR sensor covers a street point, the two vertices are connected by an edge. We call the subset of all LiDAR sensors that cover one street point $s$ the neighbors of that street point ($N_s$).

In our problem formulation, we assume that all street points must be covered, which means that for each street point, at least one of the neighboring potential LiDAR positions is equipped with an active LiDAR. At the same time, we want to minimize the total number of activated sensors.

The sensor placement problem is a version of the set cover problem, a well-known NP-hard problem \cite{KRITTER2019133}.

\subsubsection{Linear Program}

For every possible LiDAR position, we can define a binary variable $x_l$, which is one if there is a LiDAR. We aim to minimize the objective function $F_\text{obj}$, which is the sum of active sensors, while covering all street points.

\begin{align}
 \text{min} (F_\text{obj} (x_l)) = \text{min}&\sum_{V_l} x_l \\
 s.t. \quad \forall s \in V_s: \quad &\sum_{N_s} x_l \geq 1 \\
\quad \forall l \in V_l: \quad &x_l \in \{0,1\}
\end{align}

\subsubsection{QUBO Formulation}
We can formulate the problem as a quadratic unconstrained binary optimization (QUBO) problem. The general form of a QUBO problem is:

\[
\text{min} (\mathbf{x}^T Q \mathbf{x}),
\]where \( \mathbf{x} \in \{0,1\}^n \) is a vector of binary variables and \( Q \) is an \( n \times n \) matrix of real-valued coefficients. It is mathematically isomorphic to the Ising spin model with a problem Hamiltonian $\mathcal{H}$, which is the native problem representation used by current quantum annealers such as D-Wave \cite{MCGEOCH, GraphPartitioning}.
By definition, QUBO problems are unconstrained. Therefore, we define a term that imposes a penalty if the constraint is not satisfied and is zero otherwise. 

\begin{align}
 \forall s \in V_s: \quad \left(\sum_{N_s} x_l - y_s - 1 \right)^2 = 0,
\end{align}

where $y_s \geq 0$ is a slack variable ensuring that the sum can be 0 if and only if at least one LiDAR is covering street point $s$. Hence, the problem Hamiltonian is the sum of the objective function $F_\text{opt}$ and a penalty term weighted with $\alpha$.

\begin{align}
 \text{min} (\mathcal{H}) = \text{min} \left(\mathcal{H}(x_l, y_s)\right) = \nonumber \\
 \text{min} \left(\sum_{V_l} x_l + \alpha \left(\sum_{N_s} x_l - y_s - 1\right)^2\right)
\label{Hamiltonian}
\end{align}

The slack variable $y_s$ in Eq. \ref{Hamiltonian} is an integer and must be represented by binary variables for a QUBO formulation. The most common approach to achieve that is to use a one-hot encoding as formulated in \cite{Lucas}.

\begin{align}
 y_s &= \sum_{i=1}^{|N_s|} i \cdot b_i \\
 0  &= \left(1 - \sum_{k=1}^{|N_{v_i}|} b_i \right)^2 \label{onlyone}
\end{align}

Here, $y_s = (b_1, \dots, b_n)$ is represented as a binary vector of length $ n= |N_s|$ where exactly one value is 1 and the others are 0, ensured by Eq. \ref{onlyone}.

A more comprehensive encoding is using slack variables with a binary encoding:

\begin{align}
 y_s = &\sum_{i=1}^{\ceil*{\log_2|N_s|}} 2^i \cdot b_i
\end{align}

Here, $y_s$ is represented as a binary vector of length $\ceil*{\log_2|N_s|}$.

\subsection{Performance Metrics}

The selection of the right metric to measure the performance is essential to evaluate the results \cite{robustbenchmarking}. 
One suitable metric is the probability of finding the optimal solution $p_\text{opt}$ defined by the number of samples in which the optimal solution was found ($s_\text{opt}$) divided by the total number of samples ($s_\text{total}$. Another popular metric from literature is the time to solution $tts$, which is defined by the computation time required for one sample $t_s$ multiplied by the number of samples $s_{P(X)}$ that need to be done until we find the optimal solution with a certain probability $X$ \cite{robustbenchmarking}.

\begin{align}
    p_\text{opt} &= \frac{s_\text{opt}}{s_\text{total}} \\
    s_ p = s_{P(X>0.99)} &= \frac{\ln(0.01)}{\ln(1-p_\text{opt})} \\
    tts &= s_{P(X>0.99)} \cdot t_s
\end{align}

The optimal solution for the medium-scale, real-world inspired examples evaluated in this work can be found with classical approaches. It is important to note that the time for one sample $t_s$ is not representative, as the main wall clock time is spent to find an embedding and network overhead. Thus, we decided not to use $tts$ to evaluate the conducted experiments but to use $s_p$.

Another important consideration is that for larger problems, it is generally hard to find the optimal solution with quantum annealing. If $p_\text{opt} < 0.05\%$ for, e.g., 1000 samples, the results are statistically not significant. To handle this issue, we defined another metric based on a relaxed optimum, which lies within $10\%$ of the optimal value, derived from the concept of ``time to target" \cite{lubinski}. We call the relaxed optimum $p_\text{opt,rel}$ and the amount of samples to achieve it $s_{p,\text{rel}}$.

There is a special case in which no optimal solution is found during all reads. In this case, $p_\text{opt}$ is assumed to be 0 and $s_p$ is set to 5000. These assumptions are reasonable, because if one optimal solution is found in 1000 reads 
$s_p = \frac{\ln(0.01)}{\ln(1-0.001)} \approx 4603 < 5000$.


\subsection{Hyperparameters}

The D-Wave quantum annealer provides a range of tunable hyperparameters that can significantly affect the solution quality \cite{KingM14}.

It is important to consider whether hyperparameters should be tuned for each problem individually, for groups of similar problems, or globally across all problems. We chose the latter approach, as optimizing for individual problems could lead to overfitting and unrealistic results — in industrial scenarios, it is not feasible to do hyperparameter tuning for each problem. However, finding parameters that work well across all problems is more difficult and may not achieve the full potential of the solver.

\subsubsection{Penalty Weight}
The penalty terms are used to ensure valid constraints. As one more activated LiDAR increases the objective function value $F_\text{obj}$ by one, each street point not covered in a solution should be penalized stronger than that. 
Lower penalty terms can lead to infeasible solutions while higher penalty terms tend to produce worse solutions, because the couplers of the annealer struggle to handle high weights.

\subsubsection{Chain Strength}

One critical parameter is chain strength, which governs the coupling strength between physical qubits forming a logical qubit. An insufficient chain strength may lead to broken chains, causing inconsistencies in the logical representation of the problem, whereas an excessively high value leads to problems during the anneal process, as high weights can influence nearby couplers.

\subsubsection{Annealing Schedule}

The annealing schedule consists first of the annealing time, a very important parameter as the optimal solution is guaranteed for a sufficiently long annealing time. However due to noise a long annealing time can also lead to more errors, which often introduces a trade-off.

In addition to the overall annealing time, a schedule allows the user to control different annealing speeds during the process, including the ability to pause annealing for a certain time \cite{marshall2019power}.

\subsection{Decompositions}

Decomposition is a valuable tool for splitting large-scale optimization problem scenarios into feasible subproblems. Especially for quantum computing, because of hardware limitations in qubits available today, decomposition can help to embed large problem instances on quantum hardware.

\begin{figure}[h]
    \centering
    \includegraphics[width=0.99\columnwidth]{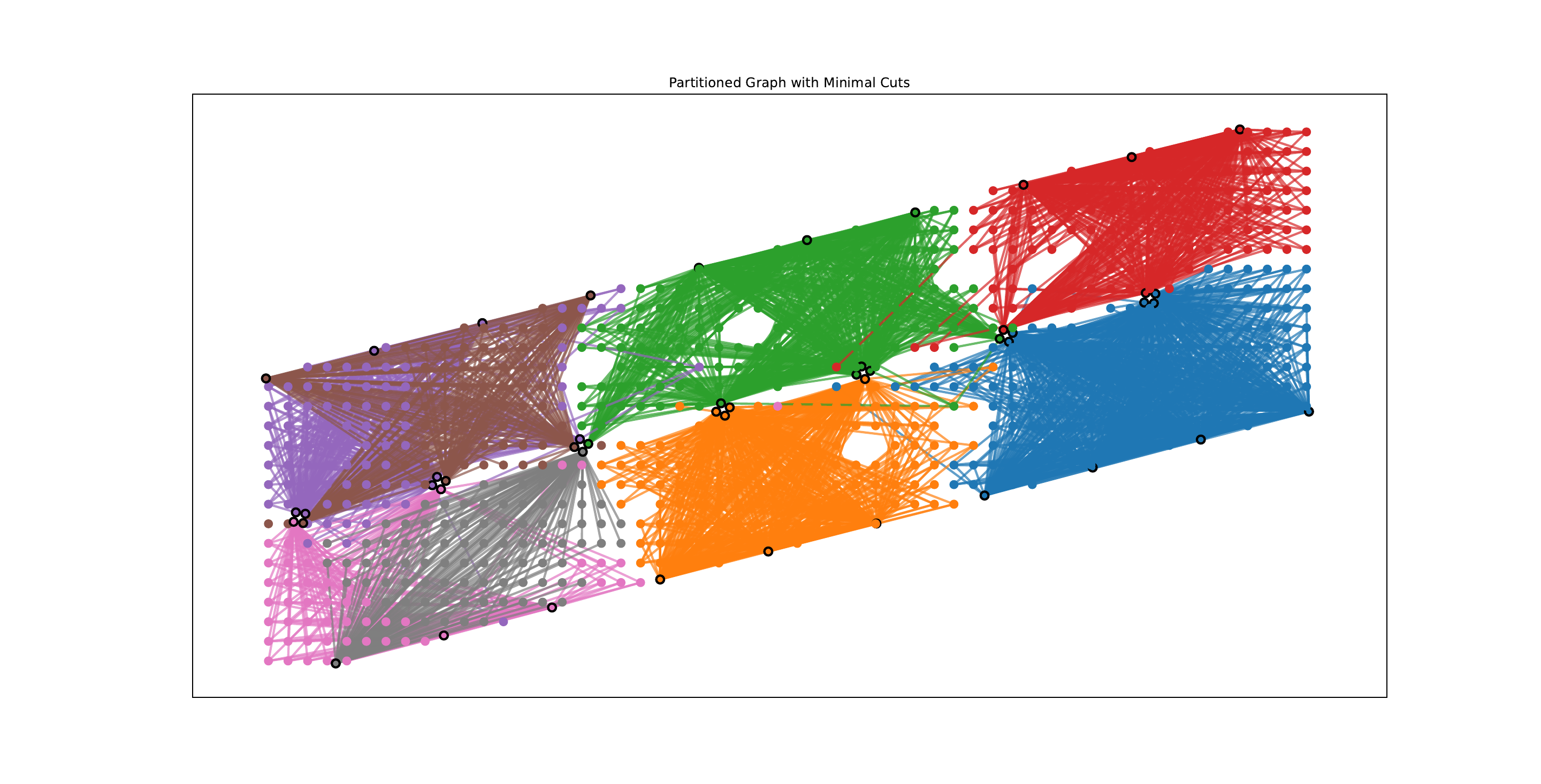}
    \caption{Spectral clustering into 8 subproblems for problem instance \textit{Real 10}.}
    \label{fig:clustering}
\end{figure}

In this work, spectral clustering is used~\cite{clustering}. This method splits a graph into subgraphs by considering nodes that are close to each other. It iteratively creates separate clusters of nodes. This algorithm prefers solutions with equally sized graphs. Fig.~\ref{fig:clustering} shows a decomposition by clustering into eight subproblems.

Further, we investigated two additional decomposition methods. Firstly, a vertical cut decomposition, where the graphs based on real-world environments can be divided intuitively into two or more parts by splitting them with straight lines. Secondly, the Kernighan-Lin bisection algorithm, a heuristic graph partition algorithm~\cite{Kernighan_Lin} separates a graph into two subgraphs while minimizing the amount of edges between the two subgraphs. However, these methods have shown no advantage over spectral clustering.


To create smaller subproblems, we use an iterative process to divide each subproblem into two new subproblems. In this way, we create $2^n$ subproblems in $n$ steps. This allows us to create sufficiently small instances even for very large problems.

It is important to note, that if the problem is feasible the combined solution of all subproblems is also feasible, since every street point within the respective partitioned areas must be covered in all subproblems, and thus in the combined problem. However, the combined solution might not be optimal even if each subproblem was solved optimally, as a sensor from one subproblem could potentially cover many street points from a neighboring area.

\section{Methodology}
\label{sec:method}

\subsection{The Problem}

In order to evaluate the performance of the quantum annealing approach studied in this work it is essential to select appropriate problem instances. Small instances can help to validate the algorithms, while larger scenarios indicate how the performance scales with the problem size.

The following scenarios are used in the experiments conducted for this work:

\subsubsection{Toy Problem Instances}
Toy problem instances are characterized by the presence of one or two walls, which serve as potential locations for the placement of LiDAR sensors, and feature one to three rows of street points situated between these walls. The horizontal dimension of each instance (referred to as ``layer") scales with the total number of potential LiDAR sensor positions and street points. The range of layers in all considered toy problem instances varies between 2 and 11. The smallest instance, referred to as \textit{Toy 1}, comprises two LiDAR sensors and two street points, resulting in a QUBO formulation with four qubits. Conversely, Fig.~\ref{toy11} illustrates the largest toy problem instance \textit{Toy 11}, which consists of 11 layers, two walls, and three rows of street points between them, thereby providing 22 potential LiDAR positions and 33 street points, culminating in a QUBO representation with 111 qubits.

\subsubsection{Real-world Inspired Instances}
\begin{figure}[h]
    \centering
    \begin{subfigure}[b]{0.4\linewidth}
        \centering
        \includegraphics[width=\linewidth]{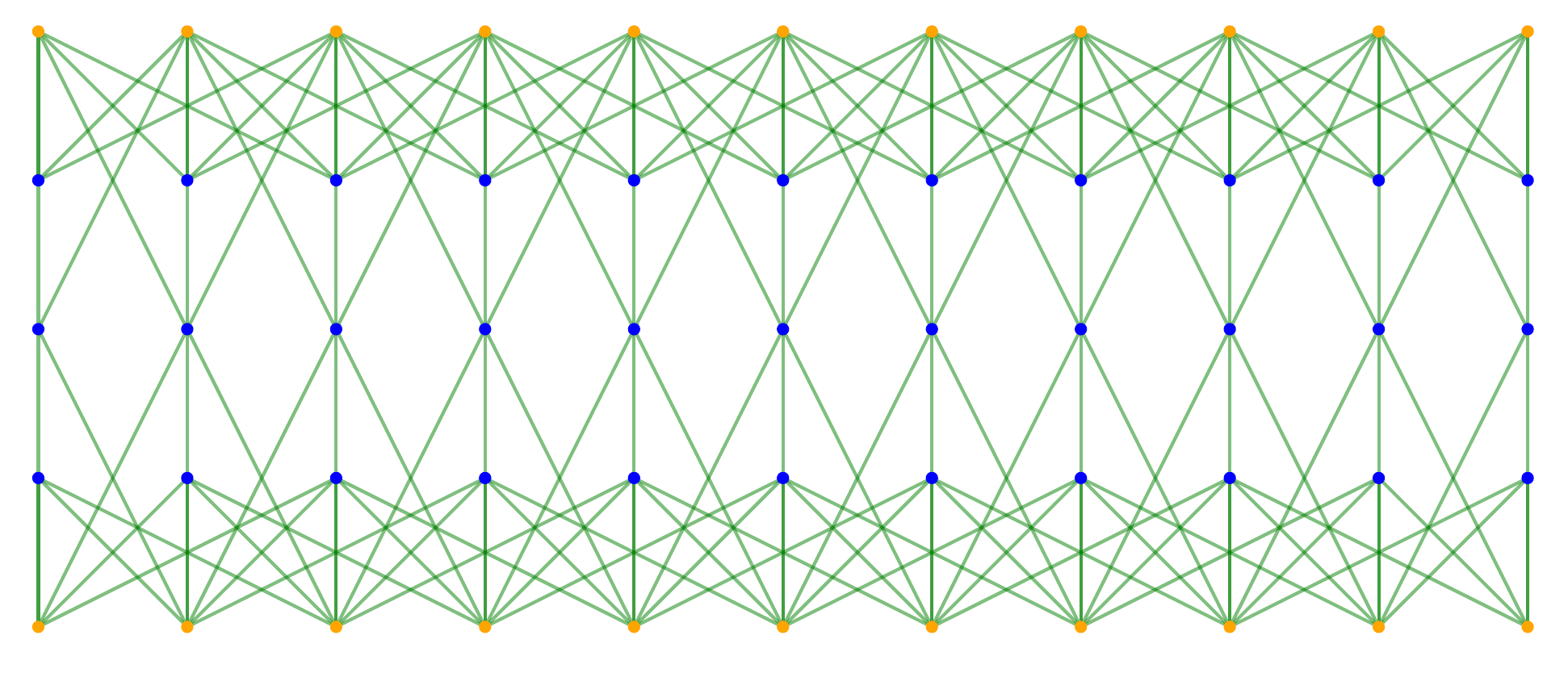}
        \caption{Problem instance \textit{Toy 11}}
        \label{toy11}
    \end{subfigure}
    \hspace{0.08\linewidth}
    \begin{subfigure}[b]{0.4\linewidth}
        \centering
        \includegraphics[width=\linewidth]{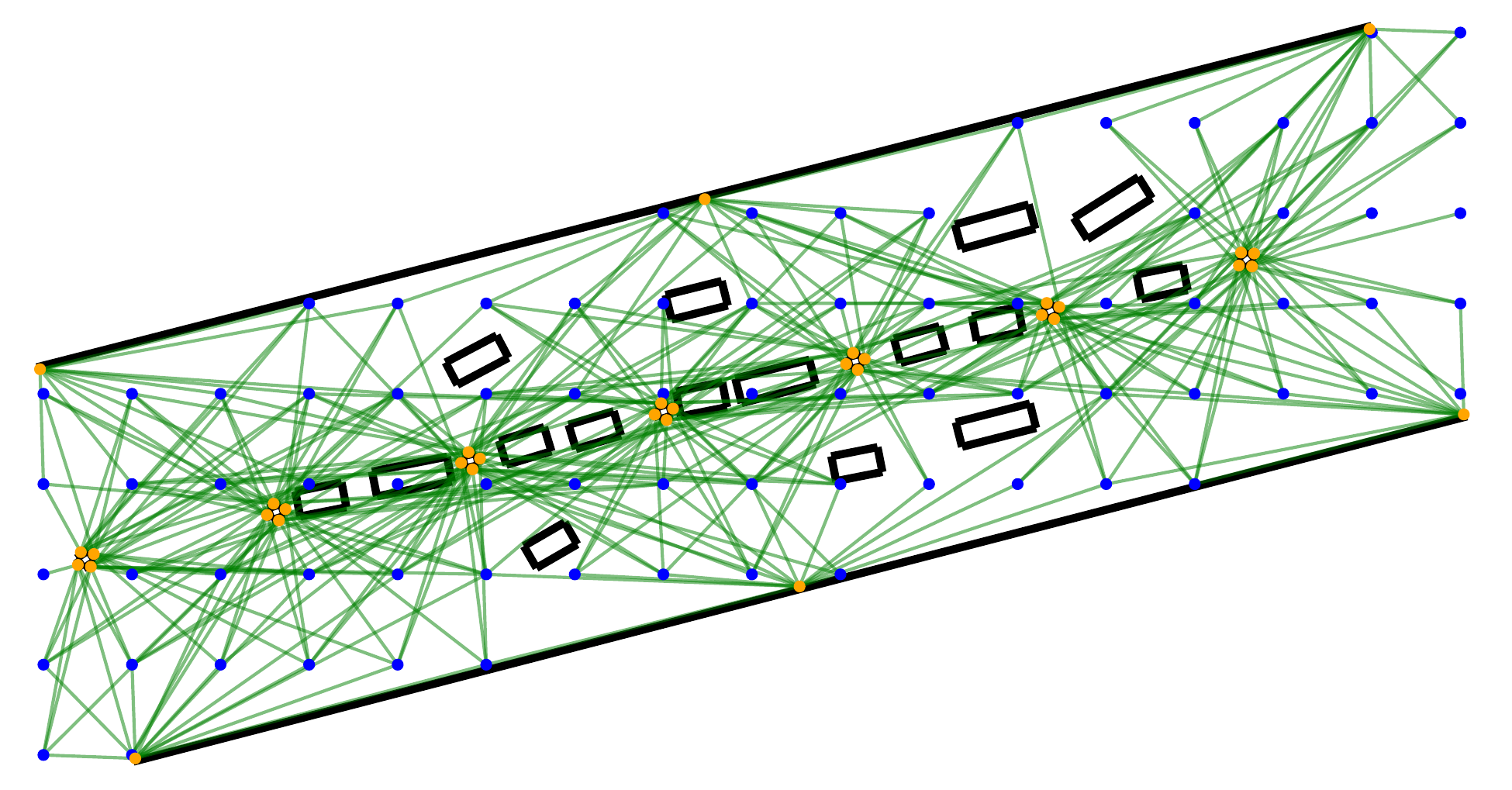}
        \caption{Problem instance \textit{Real 3}}
        \label{real3}
    \end{subfigure}
    \caption{Exemplary problem instances.}
\end{figure}

The problem instances inspired by real-world production environments that are used in this work include additional pillars between the walls, on which LiDAR sensors can also be placed. In addition, there are several obstacles that potentially block the line of sight of sensors. In these ``real-world" scenarios, the street points density is defined to be five times higher than the density of possible LiDAR positions. In Fig.~\ref{real3}, the problem instance \textit{Real 3} is depicted, in which the distance between potential LiDAR positions on the walls is $33.3$ meters, while the street points grid distance is $6.6$ meters. The smallest real-world inspired problem instance, \textit{Real 1}, includes 30 possible LiDAR positions and 18 street points, which translates to 40 qubits. while the largest problem, \textit{Real 12}, includes 52 possible LiDAR positions and 2210 street points, leading to a QUBO formulation with 4109 qubits. We can find embeddings without decomposition methods up to problem \textit{Real 6} with 879 qubits.

\subsection{Hyperparameter Optimization}

Numerous approaches for hyperparameter optimization have been documented in the literature \cite{Willsch2022, barbosa_bayesian_tuning}. Our experiments with automatic hyperparameter optimization tools revealed suboptimal performance, primarily due to significant noise in the results. In contrast, experiments with manually selected hyperparameters tend to produce more meaningful results, as human evaluators can logically assess trade-offs between parameter configurations. For instance, one parameter may increase $p_\text{opt}$ but simultaneously result in a higher incidence of infeasible solutions. Such trade-offs cannot be comprehensively captured within the objective function, but weighted and prioritized by comparing different solution metrics between various parameter setups. This paradigm allowed us to demonstrate that meaningful improvements over default settings are achievable through relatively simple and interpretable tuning strategies.

Given the numerous parameter dimensions involved in our experiments, many of which with a wide range of potential values, an exhaustive grid search is impractical. While random parameter selection is an option, it fails to adequately explore the extensive solution space. Consequently, we adopted an iterative approach, evaluating one parameter at a time and utilizing the best-performing parameter to inform the selection of the subsequent parameters. This methodology allows us to explore approximately ten possible choices for each parameter. However, a notable limitation of this approach is that it does not account for all possible combinations of parameters, which may result in the algorithm becoming trapped in local optima.

For the hyperparameter analysis, we produced 1000 samples per problem instance and parameter configuration, using ten different, precalculated embeddings for the D-Wave hardware, and calculated the average over all problem instances of the respective category (\textit{Toy} or \textit{Real}). This approach ensures a fair comparison, since embeddings have a significant impact on solution quality. Using the same set of embeddings across all parameter configurations makes the results more consistent and comparable.

The following order of parameters was chosen for the iterative optimization process: (1) QUBO encoding, (2) penalty term weights, (3) chain strength factor, and (4) annealing time.

\section{Experiments \& Results}
\label{sec:experiments}
\subsection{Experimental Setup}


Classical solutions were obtained from Gurobi \cite{gurobi} to find the optimal solution and calculate $p_\text{opt}$ and $s_p$ as well as the relaxed versions $p_\text{opt,rel}$ and $s_{p,\text{rel}}$.

For the quantum annealing experiments, we used D-Wave Advantage 4.1 and its default parameter options as benchmark \cite{advantage}.

\subsection{Hyperparameter Analysis}

For the following experiments, we tune different hyperparameters one by one and measure the performances in terms of (a) samples needed to find the optimal solution (\textit{samples}) and (b) samples needed to find a solution that is within $10\%$ of the global optimum (\textit{samples near opt}). The investigated problem instances are dissected into toy problems and real-world inspired problems (referred to as \textit{Toy} and \textit{Real} in the figures, respectively) as described in Section~\ref{sec:method}. It is important to note that both categories contain a variety of different problem sizes, which results in a high standard deviation.

\subsubsection{Encoding of the QUBO}

\begin{table}[h]
\centering
\begin{tabular}{llcc}
\toprule
\textbf{Metric} & \textbf{Category} & \textbf{Binary} & \textbf{One-hot} \\
\midrule
\textit{Samples} & Toy  & 2657 & 2882 \\
\textit{Samples near opt}  & Toy  & 1487 & 1884 \\
\textit{Samples}  & Real & 5000+ & 5000+  \\
\textit{Samples near opt}  & Real & 4005 & 4980 \\
\bottomrule
\end{tabular}
\caption{Encodings and expected number of samples for 99\% probability to include the optimum (\textit{samples} and relaxed optimum \textit{samples near opt}, i.e. within 10\% of the optimal.}
\label{tab:encoding}
\end{table}


Comparing the binary and the one-hot encodings in Table~\ref{tab:encoding}, the binary encoding produces (close to) optimal solutions with fewer samples. For real-world inspired problems, neither encoding can find the optimal solution within 1000 samples, corresponding to an expected 5000 samples for a 99\% probability of finding the optimal solution. However, in toy problem instances and in terms of finding near-optimal solutions for larger scenarios, the binary encoding performs consistently better than the one-hot encoding. The following hyperparameter analyses were therefore conducted using the binary QUBO encoding.

\subsubsection{Penalty Weight}

\begin{figure}[!h]
    \centering
    \begin{subfigure}[b]{0.48\columnwidth}
        \centering
        \includegraphics[width=\linewidth]{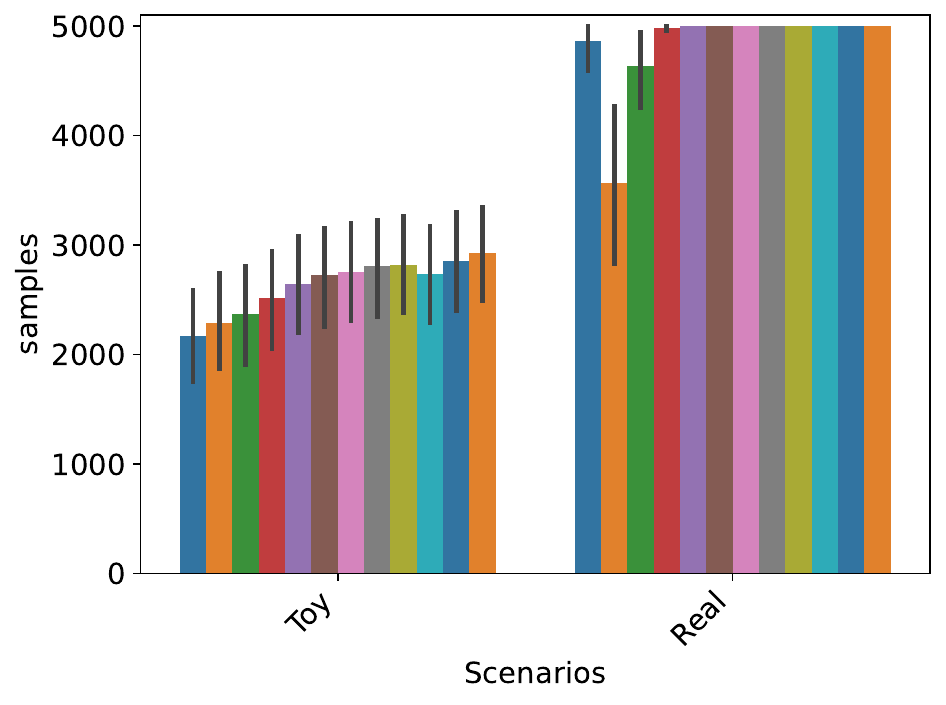}
        \caption{Samples to reach optimum.}
        \label{fig:penalty-opt}
    \end{subfigure}
    \hfill
    \begin{subfigure}[b]{0.48\columnwidth}
        \centering
        \includegraphics[width=\linewidth]{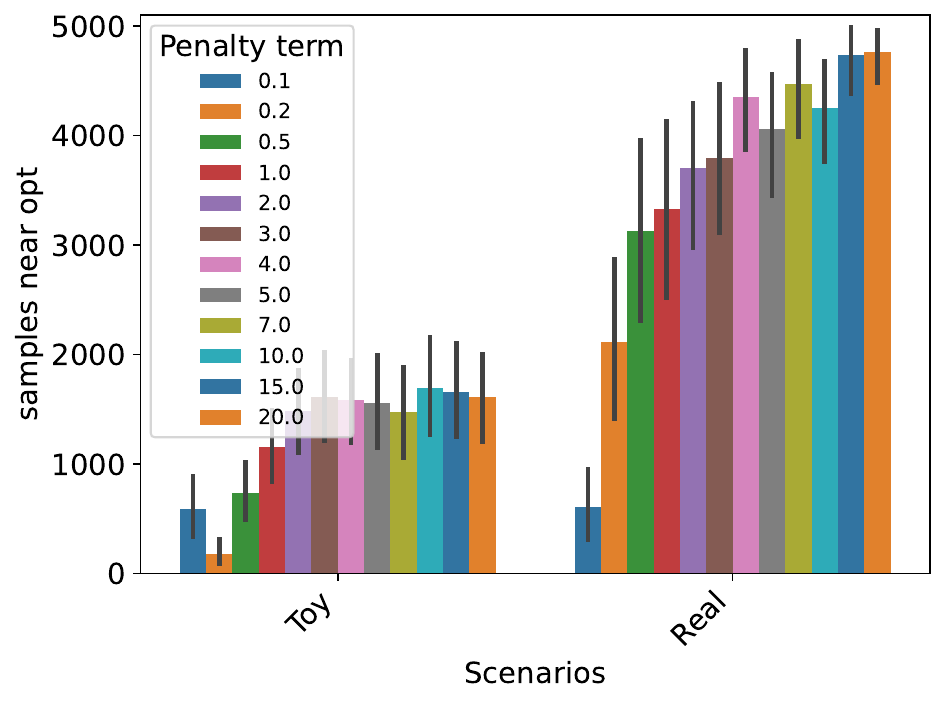}
        \caption{Samples to relaxed optimum.}
        \label{fig:penalty-relax}
    \end{subfigure}
    \caption{Penalty terms $\alpha$ and expected number of samples for 99\% probability to include the optimum (a) and relaxed optimum (b), i.e. within 10\% of the optimal. Error bars represent 95\,\% confidence intervals estimated via bootstrapping.}
\label{fig:penalty}
\end{figure}

Intuitively, in the problem Hamiltonian $\mathcal{H}$, the penalty weight $\alpha$ must be set to at least 1 to ensure that valid solutions in which all street points are covered are energetically favorable compared to solutions that violate the constraint to minimize the number of sensors used. Despite that, Fig.~\ref{fig:penalty} shows that the performance tends to be better in experiments with $\alpha<1$ compared to experiments with stronger penalized constraint violations. The best results are produced in experiments with $\alpha=0.1$ and $\alpha=0.2$. For the remaining hyperparameter analyses, we chose $\alpha=0.2$. 

\subsubsection{Chain Strength Factor}

\begin{figure}[h]
    \centering
    \begin{subfigure}[b]{0.48\columnwidth}
        \centering
        \includegraphics[width=\linewidth]{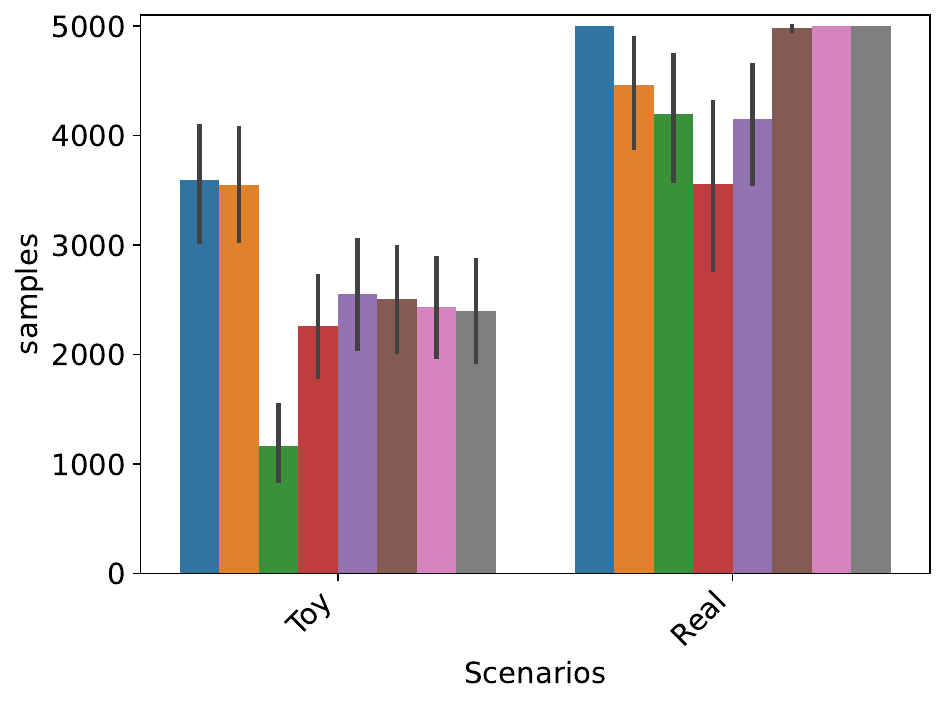}
        \caption{Samples to reach optimum.}
        \label{fig:chain-opt}
    \end{subfigure}
    \hfill
    \begin{subfigure}[b]{0.48\columnwidth}
        \centering
        \includegraphics[width=\linewidth]{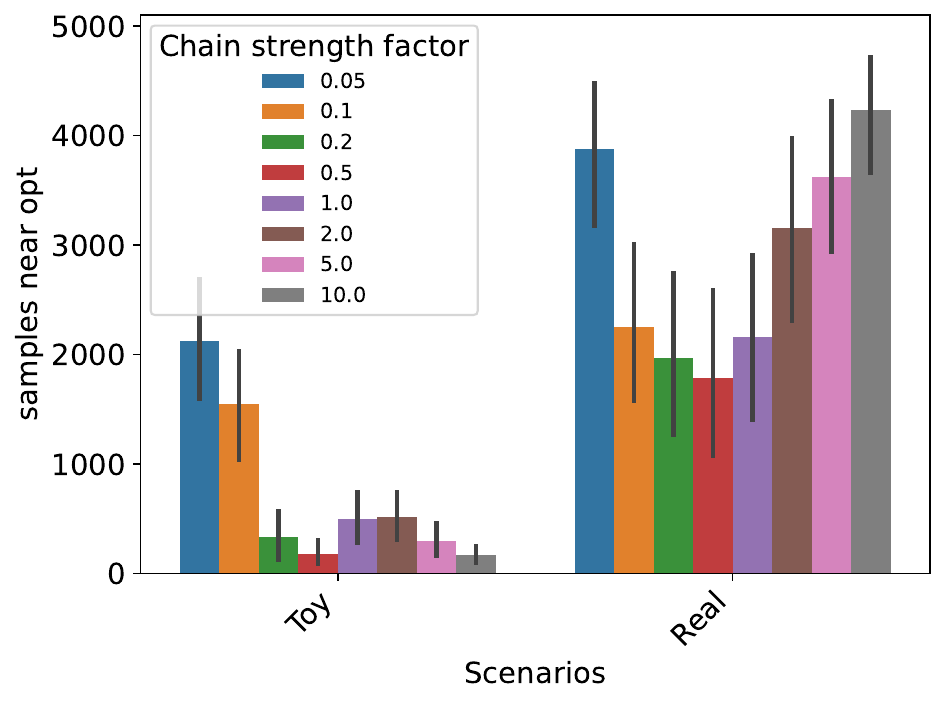}
        \caption{Samples to relaxed optimum.}
        \label{fig:chain-relax}
    \end{subfigure}
    \caption{Chain strength factors $\beta$ and expected number of samples for 99\% probability to include the optimum (a) and relaxed optimum (b), i.e. within 10\% of the optimal. Error bars represent 95\,\% confidence intervals estimated via bootstrapping.}
    \label{fig:chain}
\end{figure}

To find the optimal chain strength, we use the default chain strength provided by D-Wave (\textit{uniform torque compensation}) and multiply it with a chain strength factor $\beta$. Fig.~\ref{fig:chain} indicates that the number of samples required to identify the (close to) optimal solutions in the toy problem instances as well as real-world inspired scenarios is lowest in experiments with $\beta$ between $0.2$ and $0.5$. Due to the good performances in industry-derived problem instances, we chose $\beta=0.5$ for the rest of the hyperparameter analysis. 

\subsubsection{Annealing Time}

\begin{figure}[h]
    \centering
    \begin{subfigure}[b]{0.48\columnwidth}
        \centering
        \includegraphics[width=\linewidth]{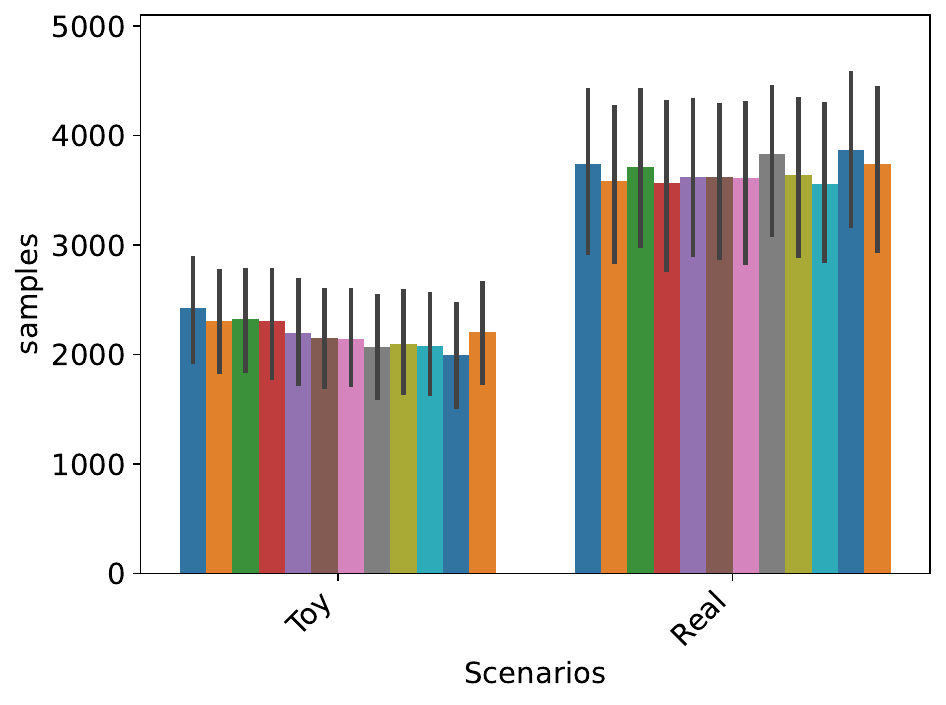}
        \caption{Samples to reach optimum.}
    \end{subfigure}
    \hfill
    \begin{subfigure}[b]{0.48\columnwidth}
        \centering
        \includegraphics[width=\linewidth]{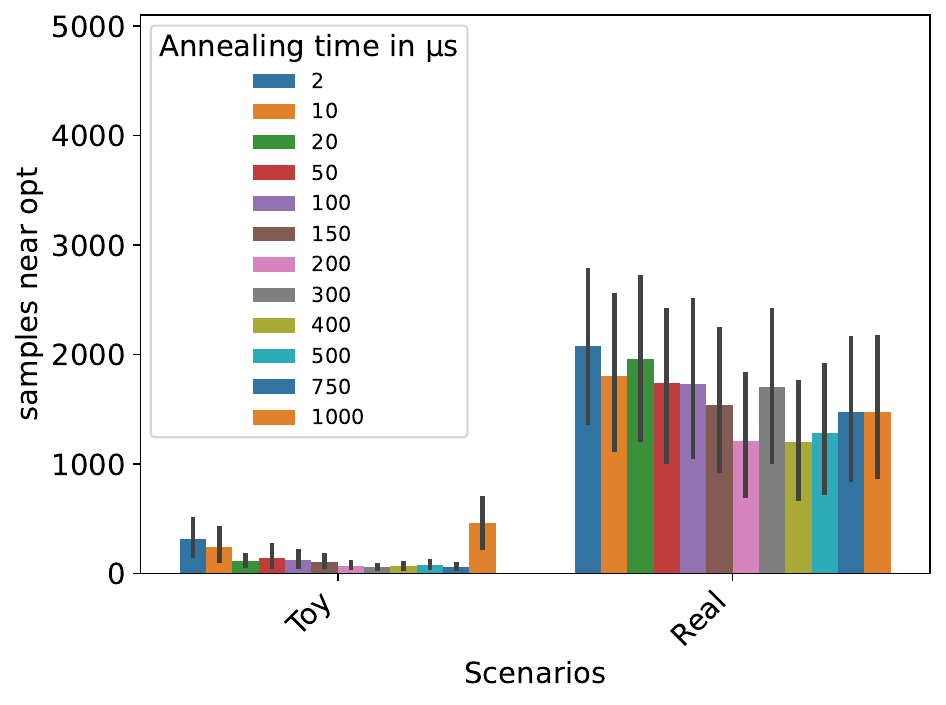}
        \caption{Samples to relaxed optimum.}
    \end{subfigure}
    \caption{Annealing times and expected number of samples for 99\% probability to include the optimum (a) and relaxed optimum (b), i.e. within 10\% of the optimal. Error bars represent 95\,\% confidence intervals estimated via bootstrapping.}
\end{figure}

In the experiments conducted in this work, the number of samples required to find a (close to) optimal solution shows little to no dependency to the annealing time parameter. There seems to be a small tendency that larger annealing times lead to fewer samples being required, albeit within the standard deviation of the results. It is important to note, that we used two batches with 500 samples each for every run, in order to not validate the maximal quantum computation time for one call set by D-Wave. We chose an annealing time of $500 \mu \text{s}$ for the following experiments. 



\subsection{Comparing Performances between the Default and Optimized Hyperparameter Configurations}

\begin{figure}[h]
    \centering
    \includegraphics[width=0.9\columnwidth]{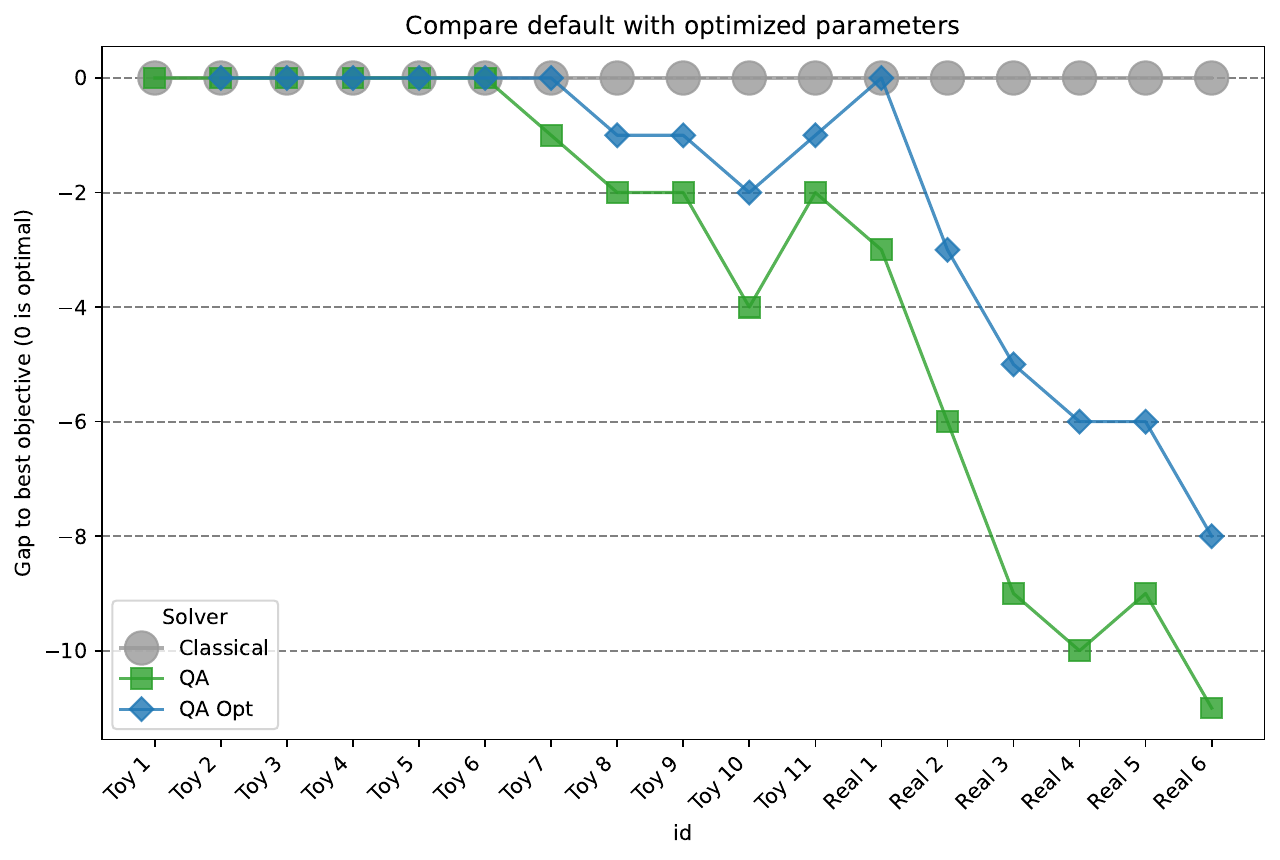}
    \caption{Comparison of default parameter setups (green) and optimized parameter setups (blue) with classically found optimum (gray) in problem instances up to \textit{Real 6}.}
    \label{fig:opt}
\end{figure}

To compare the performances of different parameterizations, we produce 1000 samples with the default hyperparameter setup provided by D-Wave and 1000 samples with the optimized parameters found in the hyperparameter analysis, select the feasible sample with the best objective value, and compare it with the global optimum of the respective problem instance found with a classical solver. It is worth noting that, due to the small penalty weight $\alpha$, this is not always the sample with the lowest value of the problem Hamiltonian $\mathcal{H}$. For example, for the smallest problem \textit{Toy 1} with two LiDAR sensors and two street points, the algorithm fails to find a feasible solution since the lowest possible value of $\mathcal{H}$ corresponds to an infeasible solution. This issue could be mitigated by increasing the penalty terms for smaller problem instances. The largest problem instance that can be solved within the 60-minute embedding time limit set by D-Wave is \textit{Real 6}.


In Fig.~\ref{fig:opt}, a significant improvement of the solution quality in experiments with optimized parameters is evident. In all instances in which default parameters can not identify the optimal solution improved solutions were obtained. In two cases (\textit{Toy 7} and \textit{Real 1}) samples with optimal solutions are produced with optimized but not with default parameters.

\subsection{Decomposition}

For this experiment, we used spectral clustering for decomposition into two and four subproblems. Since these subproblems are comparable in size to the smaller instances used during hyperparameter optimization, we consider it sufficient to reuse the previously determined parameters.


\begin{figure}[h]
    \centering
    \includegraphics[width=0.9\columnwidth]{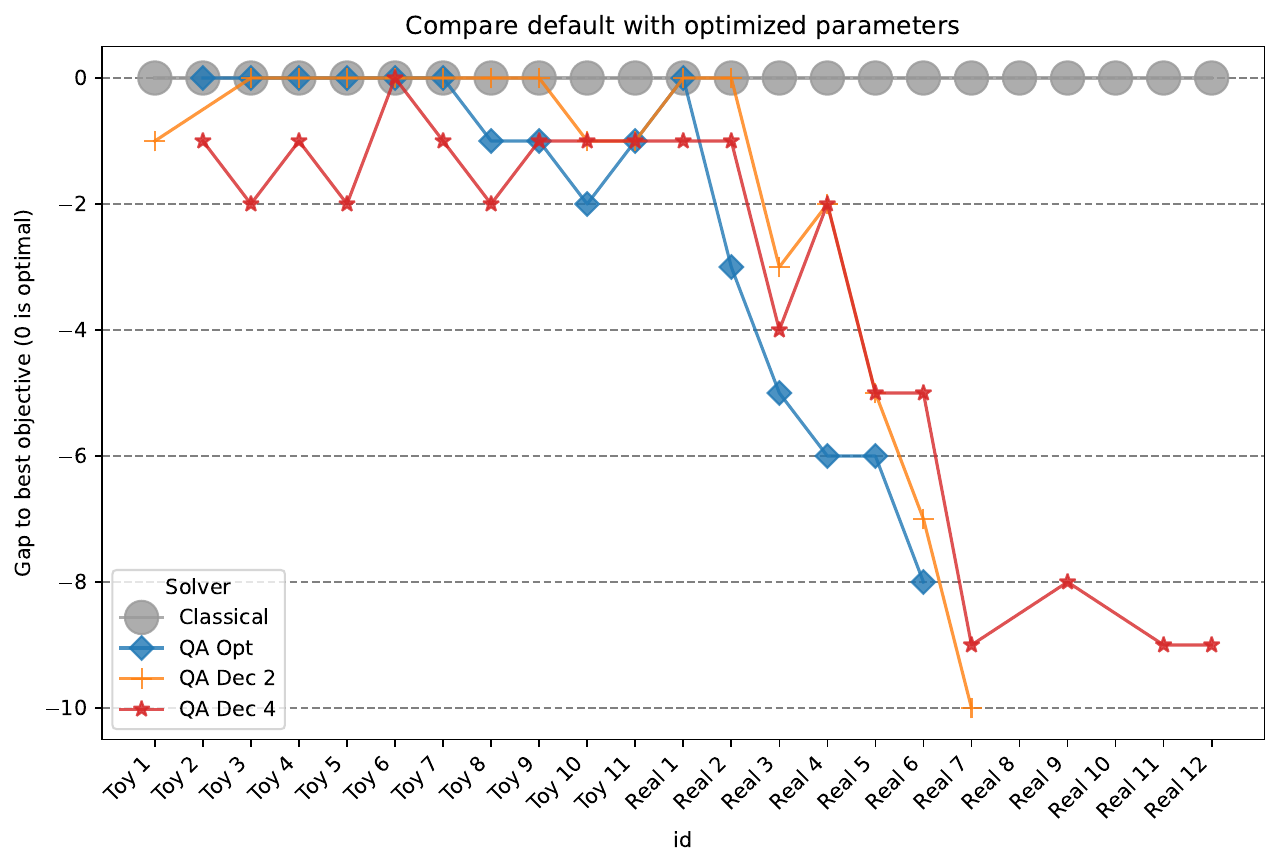}
    \caption{Comparison of optimized parameter setups without decomposition (blue), spectral clustering decompositions with two (orange) and four (red) subproblems with classically found optimum (gray) in all problem instances.}
    \label{fig:dec}
\end{figure}

Fig.~\ref{fig:dec} shows results conducted in fully composed quantum annealing experiments with optimized parameters compared with instances decomposed in two and four subproblems. 

Problem instance \textit{Toy 1} is decomposed in two subproblems with one LiDAR sensor each. In this case, two LiDAR sensors are activated to solve both subproblems. Instance \textit{Toy 2} is decomposed into subproblems with one and two LiDAR sensors, respectively, where the second subproblem can not be solved due to the small penalty value, equivalent to \textit{Toy 1} being infeasible with $\alpha<1$ as described before.

The two smallest real-world inspired problems \textit{Real 1} and \textit{Real 2} are solved optimally with quantum annealing when decomposing the problem in two subproblems. The results found in problem instances decomposed in two subproblems are better compared to the original problems in all scenarios evaluated in this work outside of the two edge-case scenarios described above.

The results obtained in experiments with problem instances decomposed into four subproblems are worse in small scenarios. Up to scenario \textit{Toy 5} this decomposition method does not perform better than \textit{QA Dec 2} in any problem instance and outperforms the composed problem formulation only once (\textit{Toy 10}).
However, by splitting larger problems into feasible parts, instances of the largest real-world inspired problem \textit{Real 12} with a total of 4109 qubits can be solved, albeit not optimally.

\section{Conclusion \& Outlook}
\label{sec:conclusion}

We evaluated quantum annealing for the optimization of LiDAR sensor configurations in automotive production environments and found that the combination of binary encoding, a low penalty weight of $\alpha=0.2$, a chain strength factor of $\beta=0.5$, and an annealing time of $500 \mu \text{s}$ demonstrates the potential to improve the solution quality compared to quantum annealing experiments with default parameters.
Decomposition techniques have proven effective in addressing larger problems closer to real-world scenarios. By decomposing the problem into two subproblems, we observed an improvement in solution quality for most cases. Further decomposition into four subproblems enabled the resolution of large problem instances with up to 4109 qubits in total through effective partition and embeddings.

We showed that hyperparameter optimization is crucial for enhancing solution quality and fully leveraging the capabilities of quantum annealing. Decomposition emerges as a robust strategy for tackling real-world inspired problem instances on current hardware, and a tailored decomposition approach should be prioritized.

Despite these findings, classical algorithms continue to significantly outperform quantum annealing. Instances derived from industrial applications are typically one order of magnitude larger than the scenarios considered in this work. In addition, potential LiDAR position density and street point density must be increased significantly to achieve a high-resolution coverage necessary for a safe and practical operation in production environments. A major challenge remains the substantial number of slack variables required to encode the derived QUBO problem, along with the additional qubits needed to embed the problem within the D-Wave hardware graph. Nevertheless, sensor positioning represents a promising application area, as we are already capable of solving comparatively large problem instances and obtaining feasible, and for smaller examples optimal, solutions.

Exploring different QUBO formulations can further enhance solution qualities, as indicated in previous studies~\cite{hristo_cover_set}. More sophisticated approaches to hyperparameter optimization are necessary, as it is impractical to fit parameters for every specific use case. A thorough benchmarking study of the sensor positioning application will be conducted in the QUARK framework~\cite{finzgar_quark_2022}, including comparisons to other quantum computing technologies like neutral-atom-based quantum computers.

To mitigate the issue of excessive slack variables, constrained mixer Hamiltonians can be employed in the quantum approximate optimization algorithm (QAOA), potentially improving quantum solutions. However, current gate model hardware is not yet mature enough for large or even medium-sized problems, limiting us to toy problems at this stage.
\section{Acknowledgments}

This paper was partially funded by the German Federal Ministry for Economic Affairs and Climate Action
through the funding program “Quantum Computing – Applications for the Industry” (QCHALLenge) based on the allowance “Development of digital technologies” (contract number: 01MQ22008).
\printbibliography

\end{document}